\documentclass[letter,12pt]{article}
\pdfoutput=1

\usepackage{jheppub}
\usepackage[T1]{fontenc}
\usepackage{slashed}
\usepackage{bbold}

\usepackage{graphicx}
\usepackage{dcolumn}
\usepackage{bm}
\usepackage{hyperref}
\usepackage{color}
\usepackage{gensymb}

\newcommand{\lp}{\left(}
\newcommand{\rp}{\right)}
\newcommand{\order}[1]{\mathcal{O}\lp#1\rp}
\newcommand{\go}{\tilde g}
\newcommand{\abs}[1]{\left| #1\right|}

\newcommand{\nc}{\newcommand}

\newcommand{\MET}{E\hspace{-2.3mm}\slash \hspace{0.2mm}_T}
\newcommand{\met}{E\hspace{-2.3mm}\slash \hspace{0.2mm}_T}
\nc{\beq}{\begin{equation}}
\nc{\eeq}{\end{equation}}
\nc{\sq}{\tilde q}

\title{The Light Gluino Gap}

\author[a,b]{Jared A.~Evans}
\author[c]{and David McKeen}

\affiliation[a]{Department of Physics, University of Illinois at Urbana-Champaign, Urbana, IL 61801, USA}
\affiliation[b]{Department of Physics, University of Cincinnati, Cincinnati, Ohio 45221, USA}
\affiliation[c]{Pittsburgh Particle Physics, Astrophysics, and Cosmology Center,\\Department of Physics and Astronomy, University of Pittsburgh, Pittsburgh, USA}

\emailAdd{jaredaevans@gmail.com}
\emailAdd{dmckeen@pitt.edu}

\abstract{A variety of experimental searches and theoretical efforts have constrained gluinos that undergo an all-hadronic decay with no missing energy, as can arise in $R$-parity violating or stealth supersymmetry.  Although the $\go\to jj$ decay is robustly excluded, there is gap in current experimental coverage for gluinos with masses between 51--76 GeV that decay into three light-flavor quarks.  In this work, we probe this gap with published multi-jet data from the UA2 experiment.  Despite setting the strongest current limit on this region, we find that UA2 data is unable to close this gap for $\go\to jjj$ decays.  In addition to this three-jet gap, we note an additional gap for all-hadronic $\go\to n~{\rm parton}$ decays with $n\geq 4$ for light gluinos (51 GeV $< m_{\go}\lesssim 300$ GeV) not covered by the current search program.}

\begin{document}
\maketitle
\flushbottom

\section{Introduction\label{sec:intro}}

 The large hadron collider (LHC) has pushed the lower bound on the gluino mass to above 1~TeV for nearly any decay path~\cite{Evans:2013jna}, and for particular decay paths, limits are above 2~TeV; see, e.g.,~\cite{Aaboud:2017vwy}.  However, there are three exceptions to this 1~TeV bound.  First, any gluino process that involves some exotic detector object requires a dedicated study and has the potential to evade the current search program if that object is not searched for explicitly (e.g., non-isolated leptons~\cite{Brust:2014gia}).  Second, gluinos that decay into all-hadronic final states exhibiting a jet $p_T$ hierarchy~\cite{Evans:2013jna}, where two jets are hard and the rest are very soft, appear more QCD-like and can have significantly weaker constraints~\cite{Khachatryan:2016xim}.  Lastly, the topic of this work, very light gluinos with all-hadronic decays, which would evade constraints by being buried within the QCD background.  
 
Independent of any production or decay modes, gluinos with $m_{\go}<51$ GeV were robustly excluded by Kaplan and Schwartz using jet thrust data from LEP~\cite{Kaplan:2008pt}, essentially a measurement of $\alpha_s$ at the $Z$ pole.   In a similar spirit, LHC measurements of the electroweak running have been used to place relatively robust constraints on new physics~\cite{Alves:2014cda}.  Although efforts have been made to extend this procedure to $\alpha_s$ information contained within the 3-jet to 2-jet ratio at the LHC~\cite{Becciolini:2014lya}, these results currently do not place robust constraints on new physics. In the CMS measurement of this ratio~\cite{Chatrchyan:2013txa}, a value for $\alpha_s(m_Z)$ is determined by varying the PDF input $\alpha_s(m_Z)$ parameter and minimizing this $\chi^2$ against the data.  This relies on SM QCD running for the profile.  Although the authors of ref.~\cite{Becciolini:2014lya} argue that the PDF profile is minimally impacted by including additional states, it is not obvious that these leading order conclusions can be extended to higher order PDFs.   While CMS uses a fixed choice of factorization and renormalization scale, the QCD hard process should in truth be sensitive to multiple scales, and the results cannot be directly used~\cite{Becciolini:2014lya}.  Moreover, very light gluinos have the potential to contribute directly to the 3-jet to 2-jet ratio, skewing any interpretation that does not contain their production and decays directly.  Of course, complementary direct limits would still be valuable even if there were a study that included all aspects necessary to indirectly exclude light gluinos.  

 All-hadronic decays of gluinos with no appreciable missing energy ($\met$) are achievable in $R$-parity violating (RPV) models~\cite{Barbier:2004ez} or stealth SUSY models~\cite{Fan:2011yu}.  Within these models, the parameters can be chosen so that the gluino would decay consistently to $n\geq2$ colored partons. While the RPV couplings require some structure to evade indirect constraints~\cite{Barbier:2004ez,Kao:2009fg}, with all other couplings small, the $\lambda''_{cds}C^cD^cS^c$ coupling in the superpotential may be $\order{1}$ without conflicting with any current precision constraints, allowing for prompt $\go\to jjj$ decays (although a complicated story for baryogenesis may be required~\cite{Graham:2012th,Cui:2013bta}).  Stealth SUSY, on the other hand,  introduces a nearly degenerate hidden sector supersymmetric singlet, $(S,\tilde S)$, that couples to QCD allowing for $\go \to g \tilde S \to g S \tilde a \to g(gg) \tilde a$ decays, where $\tilde a$ is the lightest supersymmetric particle, e.g. the gravitino or axino.  If $m_{\tilde a}<m_{\tilde S}-m_{S}\ll m_S$, the neutral, detector-stable $\tilde a$ will generically contribute missing energy far below the detector resolution, giving again a prompt $\go\to jjj$ decay. 
 
\begin{figure}[!t]
\begin{center}
\includegraphics[scale=1.]{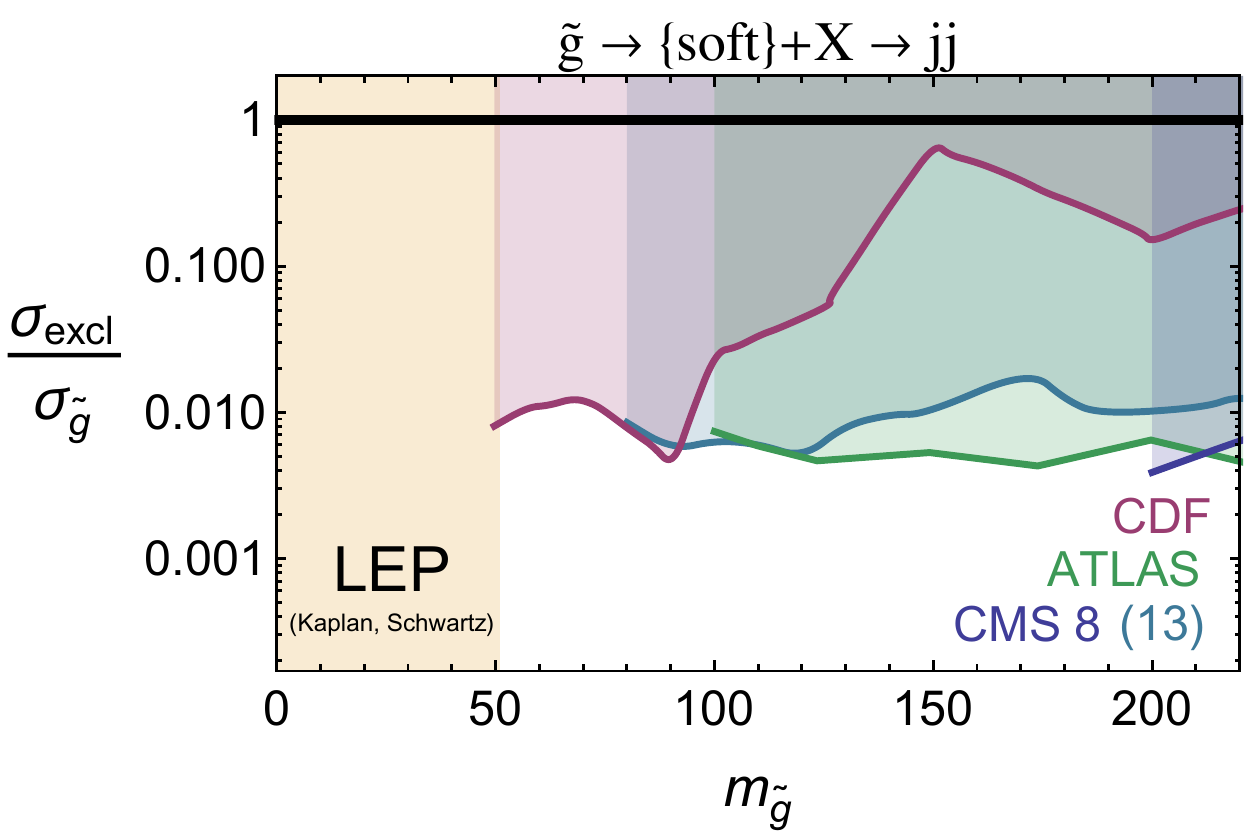}
\end{center}
\caption{Limits on stealth decays of light gluinos into (effectively) two jets, $\go\to jj\,+\,$soft.  The robust lower bound of 51~GeV from LEP jet data is in gold~\cite{Kaplan:2008pt}.  The CDF limit is in red~\cite{Aaltonen:2013hya}, the ATLAS limit is in green~\cite{Aaboud:2017nmi}, and the CMS limits from their 8 TeV direct search~\cite{Khachatryan:2014lpa} and 13 TeV substructure search~\cite{CMS:2016pkl} are in dark and light blue, respectively.}  
\label{fig:gluinolimits2j}
\end{figure}

For direct limits on all-hadronic decays, a lot of effort, from both theory and experiment, has been devoted to probing the possible all-hadronic signatures from a variety of RPV and stealth SUSY models~\cite{Fan:2012jf,Brust:2012uf,Evans:2012bf,Curtin:2012rm,Duggan:2013yna,Bai:2013xla,Evans:2013jna,Evans:2013uwa,Evans:2014gfa,Graham:2014vya,Fan:2015mxp,Monteux:2016gag,Buckley:2016kvr,Diglio:2016ynj,Beauchesne:2017jou,Dercks:2017lfq}. Although a pure $\go \to jj$ decay is not possible, if $m_{\go} -m_{\tilde S} \lesssim \order{\mbox{GeV}}$, stealth SUSY would allow for a compressed decay of $\go \to \tilde S + \{\mbox{soft}\} \to S \tilde a \to (gg) \tilde a$ to mimic a 2-jet signal.  Fortunately, experimental results at CDF~\cite{Aaltonen:2013hya}, ATLAS~\cite{Aaboud:2017nmi}, and CMS~\cite{Khachatryan:2014lpa,CMS:2016pkl} reliably close this possibility up to high masses.  The combined exclusion from these searches is summarized in figure~\ref{fig:gluinolimits2j}, where the ratio of the cross-section limit to the gluino production cross-section is shown as a function of $m_{\go}$.  Although this interpretation is simply using the efficiencies from $\tilde t\to jj$ as a function of resonance mass, the deviations introduced from switching the triplet to an octet should be $\order{1}$, at most.

  Many limits have been placed on pair-produced gluinos that each decay to three-jets by CDF~\cite{Aaltonen:2011sg}, CMS~\cite{Chatrchyan:2011cj,Chatrchyan:2012uxa,Chatrchyan:2013gia}, and ATLAS~\cite{ATLAS:2012dp,Aad:2015lea,ATLAS:2016nij}. The LHC bounds do not begin until 100~GeV~\cite{ATLAS:2012dp}, and the CDF bound begins at 76~GeV~\cite{Aaltonen:2011sg}.   Between the robust lower bound from LEP of 51~GeV and 76~GeV, where CDF becomes sensitive, there is currently a gap in the $\go\to jjj$ coverage.
  
More complicated, higher-multiplicity, all-hadronic gluino decays, such as the simple five-parton RPV decay, $\go \to jj\tilde\chi^0 \to jj(jjj)$, or, more generally, any $n\geq 4$-parton decays that can be achieved in stealth SUSY or hidden valley models~\cite{Strassler:2006im,Strassler:2006qa,Evans:2013jna}, have been excluded by black hole~\cite{Chatrchyan:2013xva} and all-hadronic~\cite{Aad:2015lea} LHC searches for $m_{\go}\gtrsim300$~GeV~\cite{Evans:2013jna}.  Below $\sim300$~GeV limits weaken, as the significant boost required for the event to pass the trigger causes the decay products to be too collimated to be resolved as separate jets.

In this work, we confront one facet of the light gluino gap, namely the RPV decay of $\go \to jjj$ within the 51--76~GeV band, by examining multi-jet data from UA2~\cite{Alitti:1991rd}.  In the next section, we will describe the UA2 detector and our UA2 simulator.  In order to validate this digital simulacrum, we reproduce the efficiencies from the only hadronic SUSY search at UA2~\cite{Alitti:1989ux}.  We then describe how we reinterpret the multi-jet measurement to place limits on light, $R$-parity violating gluinos.  While these results are close to exclusion, they are unable to close this gap for gluinos.  We conclude by discussing how the light gluino gap could be probed with targeted substructure-based LHC searches, likely closing the gap not only for three-jet decays, but also for $n\geq 4$-parton decays.

  \section{The UA2 Experiment\label{sec:UA2}} 
  
   UA2 was one of the two primary experiments at the CERN $p\bar p$ SPS collider.  Together with UA1, the collaboration co-discovered the $W$~\cite{Banner:1983jy} and $Z$ bosons~\cite{Bagnaia:1983zx} (see~\cite{Jakobs:1994fc} for more history).  Years later, when the two studies~\cite{Alitti:1989ux,Alitti:1991rd} that are the focus of this work were performed, the UA2 detector had collected $\sim 7.5$pb$^{-1}$ of $\sqrt{s}=630$~GeV data, covered nearly $4\pi$ of solid angle, displayed exceptional calorimetry, and possessed no magnetic field.  
 
 UA2 consisted of a central calorimeter covering $\abs{\eta}<1$ with cells of polar and azimuthal widths of $\Delta\theta =10^{\degree}$ and $\Delta\phi =15^{\degree}$, respectively.  The two end cap calorimeters covered $1<\abs{\eta}<3$, with cell widths of $\Delta\eta =0.2$ and $\Delta\phi =15^{\degree}$, out to $\abs \eta =2.2$ where the final cells had $\Delta\phi =30^{\degree}$ and $2.2<\abs\eta<2.5$ and $2.5<\abs\eta<3.0$.   The cells contained multilayer lead-scintillator and iron-scintillator sandwiches for electromagnetic and hadronic calorimetry, respectively.  The calorimeter exhibited an energy resolution $\sigma_E\sim 0.32 \lp\frac{E}{\mbox{\scriptsize GeV}}\rp^{\frac34}$~GeV, which we apply to the cells with a truncated gaussian about the truth-level value determined by the showering in Pythia 8~\cite{Sjostrand:2014zea}.  A muon would deposit up to 1.5~GeV of energy during its minimally ionizing traverse, with any remaining energy appearing as $\met$.  As our signal of interest does not contain electrons at appreciable rates, energy from these particles is lumped into jets for simplicity.  
 
After cell contents are recorded by the the detector, jets are formed by dropping all cells with $E<400$ MeV and combining adjacent cells above this energy threshold until no cells remain.  Unlike in some other UA2 results~\cite{Jakobs:1994fc}, cell clusters are not split in either search due to the presence of energetic valleys~\cite{KMeier}.   The energy and size of the jet are defined to be $E=\sum_i^{\rm cells} E_i$, and 
 \beq
 R=\sqrt{R_\phi^2+R_\theta^2} \mbox{ where } R_x = \sum_i^{\rm cells} R_{x,i} \frac{E_i}{E},
 \label{eq:R}
 \eeq
 where $R_{\phi(\theta),i}$ is the angle corresponding to the center of the cell.    In standard fashion, missing transverse energy is determined by performing a vector-like sum of the calorimeter cells, and projecting onto the $x-y$ plane. 
 
  \begin{figure}[!t]
\begin{center}
\includegraphics[scale=1.3]{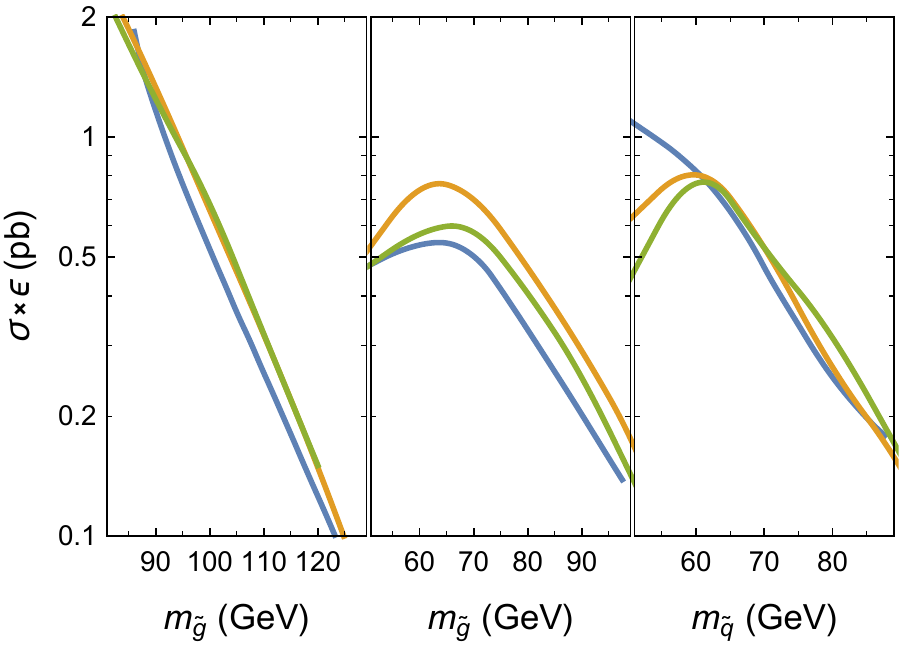}
\put(-250,220){\footnotesize$m_{\sq}=m_{\go}$}
\put(-175,220){\footnotesize$m_{\sq}=200$ GeV}
\put(-75,220){\footnotesize$m_{\go}=10^5$ GeV}
\put(-295,55){\bf\footnotesize(a)}
\put(-196,55){\bf\footnotesize(b)}
\put(-96,55){\bf\footnotesize(c)}
\end{center}
\caption{Recast of the UA2 SUSY search~\cite{Alitti:1989ux}.  The excluded effective cross-sections are shown as functions of (a) $m_{\go}=m_{\sq}$, (b) $m_{\go}$ with $m_{\sq}=200$~GeV, and (c) $m_{\sq}$ with $m_{\go}=10^5$~GeV.  In blue are the curves from UA2.  Our simulation rescaled to their cross-section is shown in orange.  In green, we show our simulation with an FSR enhancement (obtained by increasing $\alpha_s(m_Z)$ to 0.25) to better mimic the aggressive FSR profile used in the UA2 study.}  
\label{fig:SUSYeff}
\end{figure}
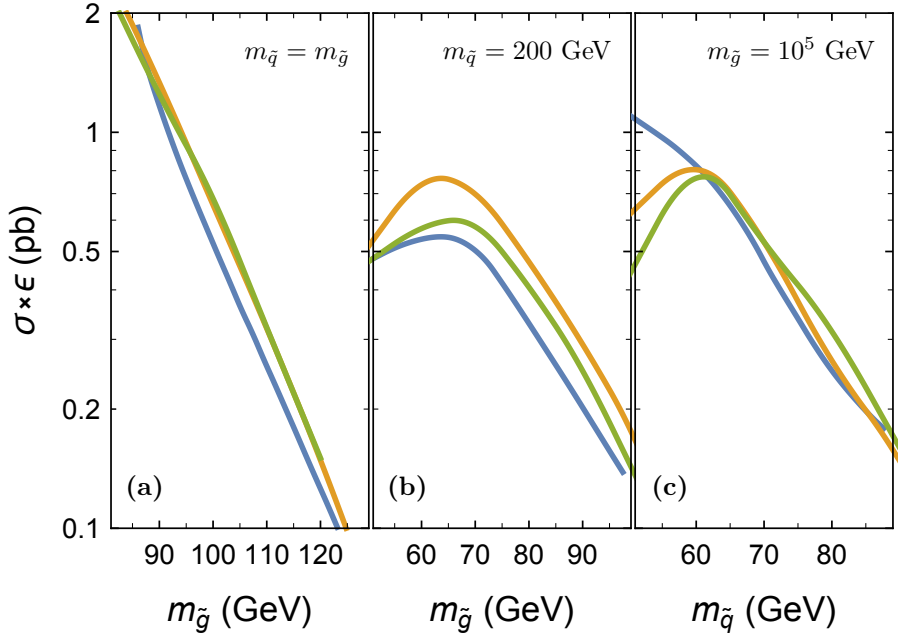
 
 In order to validate our simulation of the UA2 detector, we recast the UA2 search for squarks and gluinos~\cite{Alitti:1989ux} using their benchmark models.  The search requires at least two central jets with $E_{T,1(2)}>25 (15)$~GeV and $\abs{\eta}<0.85$ that do not behave like QCD dijet events, and $\MET>40$~GeV not aligned with jets in the event.  Event samples are generated and showered in Pythia 8~\cite{Sjostrand:2014zea}.  As many of the simulation details provided by UA2 are antiquated, we focus on the efficiency between our recast and their model.   These results are presented in figure~\ref{fig:SUSYeff}, using data from figure~4 of ref.~\cite{Alitti:1989ux} for three benchmark mass curves:  $m_{\go}=m_{\sq}$, $m_{\sq}=200$~GeV,  and $m_{\go}=10^5$ GeV (in all cases $m_{\tilde\chi^0}\approx 0$).   Although we were unable to reproduce the quoted values for their SUSY cross-sections using any tools at our disposal, we rescale our cross-section values to match the values presented in Table 1 of~\cite{Alitti:1989ux} via
 \beq
 \sigma_{\rm use}(200, m_{\go})  = \frac{\sigma_{\rm UA2}(200,80)}{\sigma_{\rm sim}(200,80)} \sigma_{\rm sim}(200, m_{\go}),
 \eeq
and similarly for the two other benchmark lines in the study.  While we do not expect this approximation to be reliable far from the central point, we believe it is sufficient for determining the reliability of our simulation.   In Pythia, we disable  multiple parton interactions, initial state radiation, and final state radiation (FSR) in process, but enable FSR in resonances to reflect their treatment of the Monte Carlo SUSY model. The UA2 result is shown in blue, with our simulation in orange.  Overall, the agreement is rather good, with the exceptions of the 50 GeV squark point (in figure~\ref{fig:SUSYeff}c), where the efficiency is very low, and in the entirely gluino sample (figure~\ref{fig:SUSYeff}b) where disagreement is about 40\% for all mass choices.   In this latter case, we believe this is due to a disagreement between FSR modeling in modern Monte Carlo tools and the fragmentation scheme used by UA2~\cite{Field:1977fa}, with the latter being overly aggressive. To support our suspicions, we reran our simulation with $\alpha_s(m_Z)$ set to the highest value allowed in Pythia (0.25) modifying the efficiencies resulting in the green curve in figure~\ref{fig:SUSYeff}.  This value more accurately reflects the gluino case, with only small changes to the other cases.  Further refinements to our archaeological efforts would not be valuable, as our simulation is mimicking the UA2 detector to the precision we require.    In the main study of this paper, we will use the default showering settings, as the modern understanding of $p\bar p$ interactions will better reflect the actual environment in SPS collisions.

   \section{Limits on gluinos from multi-jets\label{sec:limit}} 
   
   The UA2 search of interest is a measurement of multi-jet events and search for double-parton scattering~\cite{Alitti:1991rd}.  The search region of interest in this study requires six jet clusters (as defined in the previous section), each with $E_{T,j}>15$ GeV and $\abs{\eta}<2.0$, at least 40\% of the total transverse energy in the calorimeter to be contained within these six jets, the average cluster radius of the jets, as defined in (\ref{eq:R}), to be less than $R_{\rm av}<0.349$, no additional jet clusters with $E_{T,j}>10$ GeV, and small missing energy, $\MET<20$ GeV. Within the search region, UA2 observed only 7 of these six-jet events with their 7.6 pb$^{-1}$ of data.  
   
   \begin{figure}[!t]
\begin{center}
\includegraphics[scale=1.]{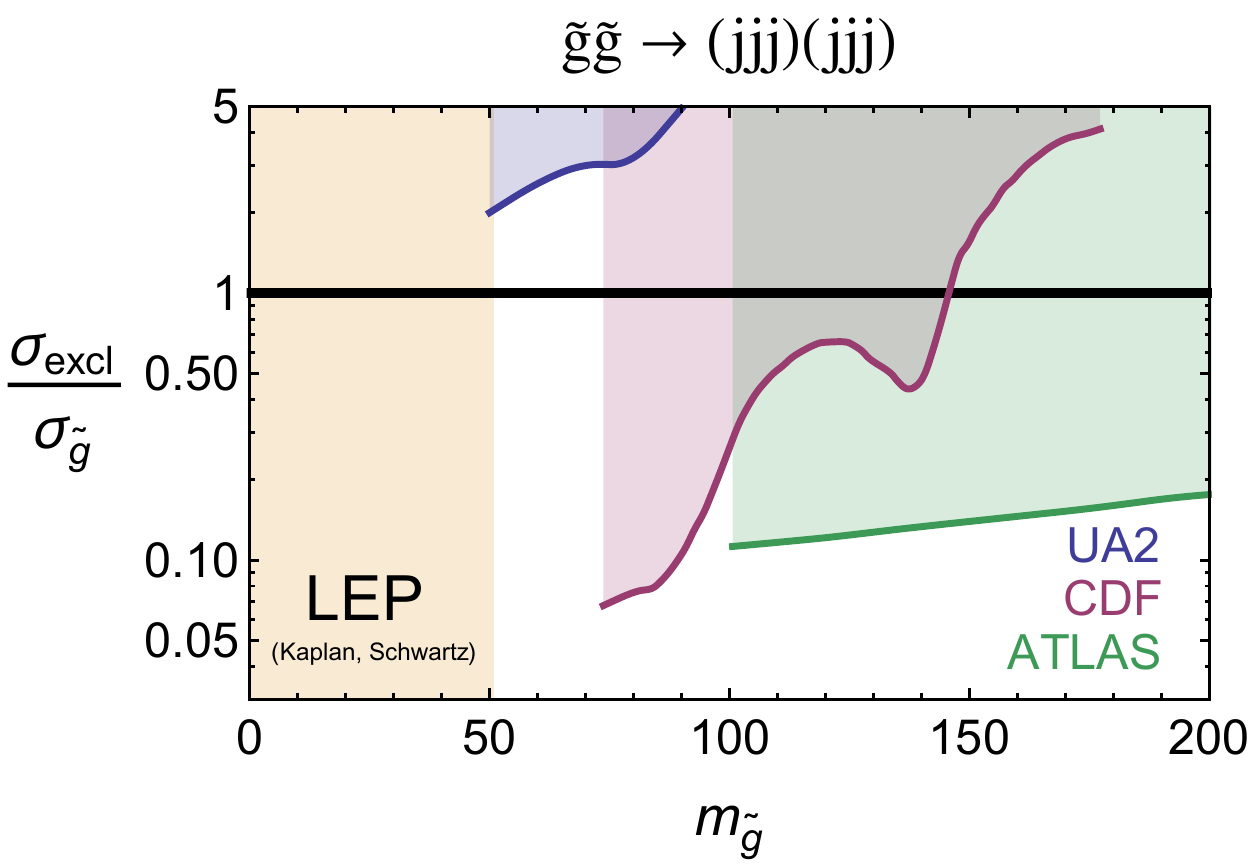}
\end{center}
\caption{ Limits on RPV decays of light gluinos, $\go\to jjj$.  The recast performed in this work of the UA2 study~\cite{Alitti:1991rd} is in blue, the CDF limit is in red~\cite{Aaltonen:2011sg}, the lowest LHC constraint from ATLAS is in green~\cite{ATLAS:2012dp}, and the robust lower bound of 51 GeV from LEP jet data is in gold~\cite{Kaplan:2008pt}.}  
\label{fig:gluinolimits}
\end{figure}
   
  Although this study is intended as a measurement of standard model properties, the signature from prompt, all-hadronic, $R$-parity violating decays of gluino pairs into three quarks, $\go\go\to(jjj)(jjj)$, can actually overwhelm the QCD background so much that $p\bar p\to \go\go\to(jjj)(jjj)$ events alone would produce many more than the seven events UA2 observed in their signal region.  We produce our $p\bar p\to \go\go\to(jjj)(jjj)$ signal events in MadGraph 5~\cite{Alwall:2014hca}, shower in Pythia 8~\cite{Sjostrand:2014zea}, and process through our custom UA2 simulator (detailed in the previous section) before applying the specific cuts used in this search.  The SPS gluino cross-sections were determined with Prospino2.1~\cite{Beenakker:1996ch,Beenakker:1996ed} (modified slightly to treat SPS energies).    We do not match additional jets, which is justified as hard radiation from the parton shower providing more than 15 GeV, or $\sim2\%$ of the total collision energy would be extremely rare.
  
  While there are several known sources of uncertainty, including the luminosity, our Monte Carlo statistics, and the PDF, the largest by far is our UA2 modeling uncertainty.   To this end, we conservatively assign a 33\% uncertainty on our signal for all choice of gluino masses, implying that signals yielding more than $N_{95,\rm excl}=19.6$ events ($\sigma_{\rm excl}=2.58$ pb) are excluded. In figure~\ref{fig:gluinolimits}, we show the cross-section as a function of gluino mass excluded by this search in blue, normalized to the gluino production cross-section value at that mass. We find that  UA2 constrains the gluino production cross-section for $R$-parity violating decays to within 2-3 times the expected values across the entire light gluino to three-jet gap.

   \section{Discussion\label{sec:discussion}} 
   
    In this work, we illustrated that six-jet event data from UA2 constrains the allowed cross-section for all-hadronic, three-body decays of gluinos in $R$-parity violating SUSY with masses from 51--76 GeV, between the sensitivity of LEP and CDF.   While unable to close the light gluino to three-jet gap, the UA2 data does exlcude gluino production cross-sections that are a factor of 2--3 larger than the expected values, setting what is presently the strongest bound within this region. 

As discussed in the introduction, high-multiplicity, all-hadronic gluino decays, $\go\to n~{\rm  partons}$ with $n\geq 4$, have been excluded for $m_{\go}\gtrsim300~\rm GeV$ by LHC searches~\cite{Chatrchyan:2013xva,Aad:2015lea,Evans:2013jna}. However, these states are not currently robustly constrained at lower masses, as the decay products become too collimated for the existing multi-jet searches to have sensitivity.  Further LHC studies, such as dedicated substructure-based searches within dijet events, have the potential to close this gap for all $n$-parton decays.  Jet substructure techniques (for a recent review and original references, see~\cite{Larkoski:2017jix}) have been successfully used to reduce large QCD backgrounds in a variety of searches.  These method have, for instance, allowed CMS to search for new bosons with masses as low as $50~\rm GeV$ that decay into dijets~\cite{Sirunyan:2017nvi}, and to search for paired dijet resonances with masses as low at 80 GeV~\cite{CMS:2016pkl} (displayed in figure~\ref{fig:gluinolimits2j}).  In addition to probing $\go\to n~{\rm  partons}$, such substructure-based searches would have the potential to constrain light squark multi-jet signatures, such as $\sq \to j \tilde\chi^0\to j(jjj)$ in RPV or $\sq \to j\tilde S \to j(gg) \tilde a$ in stealth SUSY, or all-hadronic decays of non-SUSY particles such as colorons or axigluons, see e.g.~\cite{Han:2010rf,Bai:2011mr,Gresham:2012kv}.  Direct probes of light hadronically-decaying particles at the LHC are desirable, due the relatively small theoretical and modeling uncertainties, and timely, as rising trigger thresholds will make accessing these regions more difficult.

\begin{acknowledgments}
  We thank J.~P.~Chou, S.~D.~Ellis, G.~Gollin, D.~Goncalves, E.~Halkaidakis, J.~Incandela, A.~Larkoski, S.~Mukhopadhyay, A.~Nelson, H.~Russell, D.~Shih, and S.~Thomas for useful discussions.  We thank  J.~Shelton for collaboration in the early stages of this work.  We are especially grateful to Karlheinz Meier for providing specific details concerning the jet algorithms at UA2.  The work of JAE is supported in part by DE-SC0015655. The work of DM is supported by PITT PACC through the Samuel P. Langley Fellowship.
  \end{acknowledgments}
  

\bibliographystyle{JHEP}
\bibliography{UA2Gluinos}


\end{document}